\documentclass[aps,prl,twocolumn,floats,superscriptaddress]{revtex4-2}
\usepackage{graphicx}
\usepackage{dcolumn}
\usepackage{bm}
\usepackage{amsmath}
\usepackage{color}
\usepackage{amssymb}
\usepackage{ulem}
\usepackage{xcolor}
\usepackage{subfigure}

\usepackage{amssymb}
\usepackage{latexsym}
\usepackage{epsfig}
\usepackage{psfrag}

\usepackage{amsmath,amsfonts,amssymb,bm,ulem}
\usepackage{color}
\usepackage{float}

\graphicspath{{figs/}}

\newcommand{\be}{\begin{equation}}
\newcommand{\ee}{\end{equation}}
\newcommand{\bea}{\begin{eqnarray}}
\newcommand{\eea}{\end{eqnarray}}
\newcommand{\beq}{\begin{equation}}
\newcommand{\eeq}{\end{equation}}

\begin{document}
\author{Nikolay Prokof'ev}
\affiliation{Department of Physics, University of Massachusetts, Amherst, MA 01003, USA}

\author{Ilya Esterlis}
\affiliation{Department of Physics, University of Wisconsin-Madison, Madison, WI 53706-1390, USA}

\author{Artem Abanov}
\affiliation{Department of Physics,  Texas A\&M University, College Station,  TX 77843, USA}

\author{Andrey Chubukov}
\affiliation{Department of Physics, University of Minnesota, Minneapolis, MN 55455, USA}

\title{Limits of Validity for Migdal-Eliashberg Theory: role of Polarons and Bipolarons }
\begin{abstract}
It is widely  believed that in the adiabatic limit the Fermi liquid  state of an electron-phonon system,  described by Migdal-Eliashberg theory,  remains  stable  until the dressed phonon softens.  Our   variational and analytic  analysis  of the prototypical  Holstein model  shows that,  in a wide range of fillings both in 3D and 2D,
a polaronic/bipolaronic state emerges  before phonon softening;  at  small filling  in 3D  this happens already
at weak coupling.  We show that a polaronic/bipolaronic state  emerges, upon increasing coupling,  via an intermediate pseudogap-type mixed state, in which some fermions retain Fermi liquid behavior, yet Luttinger's theorem is violated.
At even larger couplings the density of states gradually approaches its form in the atomic limit.
\end{abstract}

\maketitle
{\it {Introduction}}~~~
Migdal-Eliashberg theory (MET) is an established paradigm for describing Fermi liquid
(FL) and superconducting states in a system of electrons interacting with phonons
\cite{Migdal,Eliashberg,AGD}. At weak coupling, MET is synonymous with 
perturbation theory,  with the advantage that no artificial high-energy cutoff is required;
e.g., superconducting $T_c$ can be computed with
 the  prefactor~\cite{Dolgov_2005,*Wang_2013,Andrey_review,Mirabi_2020,marsiglio2020eliashberg,Kiessling2025,*Kiessling2025_1,*Gnezdilov_2025}.
At strong coupling, MET states that, in the adiabatic limit where the Debye frequency, $\omega_0$, is much smaller than the Fermi energy, $E_F$,  one can proceed with the self-consistent one-loop approximation despite large enhancement of the
fermionic mass because vertex corrections remain small in $\gamma=\omega_0/E_F \ll 1$; see Ref.~\cite{Migdal}.
According to MET, such a regime develops at coupling $\lambda_0$ below
the one at which the dressed phonon frequency vanishes at some momentum.
The existence of the strong coupling MET regime has been questioned recently~\cite{Yuzbashyan,*Yuzbashyan_1,
*Yuzbashyan_3},
yet the validity of MET at smaller couplings,  and especially
in the  weak coupling  limit,  has never been questioned.

The textbook model for MET consists of 3D electrons with parabolic dispersion, an optical mode with
dispersion $\omega_0({\mathbf q})$, and density-displacement
electron-phonon (e-ph) interaction $g$ \cite{AGD}:
\begin{equation}
H_{\rm int} = \sum_{{\mathbf q} } \, \frac{g}{\sqrt{2\omega_0({\mathbf q})}} \,
\left[ \rho_{\mathbf q} b_{\mathbf q} + h.c. \right] .
\label{Hint}
\end{equation}
Here $\rho_{\mathbf q}= \sum_k c^\dagger_{k+q} c_k$ is the Fourier transform of the total electron density
and $b_{\mathbf q}$ is the phonon annihilation operator
\footnote{ In general, the interaction $g$ can also depend on $q$,
but this dependence does not critically affect our findings and we neglect it.}.
For this model, the dimensionless coupling is
\begin{equation}
\lambda_0 = N_F \left\langle \frac{g^2}{\omega^2_0({\mathbf k}-{\mathbf k}')} \right\rangle _{FS} ,
\label{l0}
\end{equation}
where $N_F$ is the density of states at the Fermi energy and
$\left\langle \dots \right\rangle _{FS}$
stands for averaging over ${\mathbf k}$ and ${\mathbf k}'$
on the Fermi surface (FS). Within MET,
the phonon frequency is renormalized into
$\omega^2_r({\mathbf q}) = \omega^2_0({\mathbf q}) -g^2\Pi_{st}({\mathbf q})$,
where the static polarization $\Pi_{st}({\mathbf q})$ is a convolution of
two dressed Green's functions, which in the
adiabatic limit can  be approximated by the bare ones.
\begin{figure}[htbp]
	\includegraphics[width=0.4\linewidth]{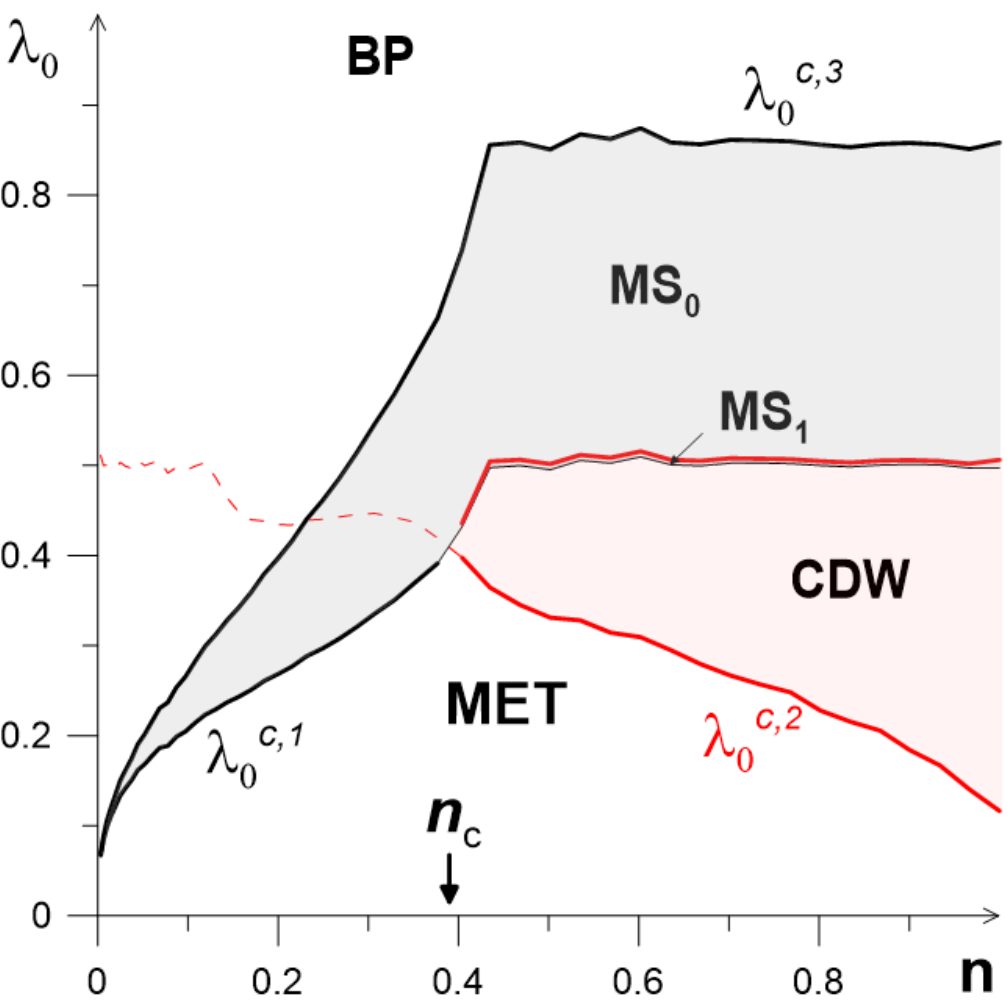}
    \includegraphics[width=0.4\linewidth]{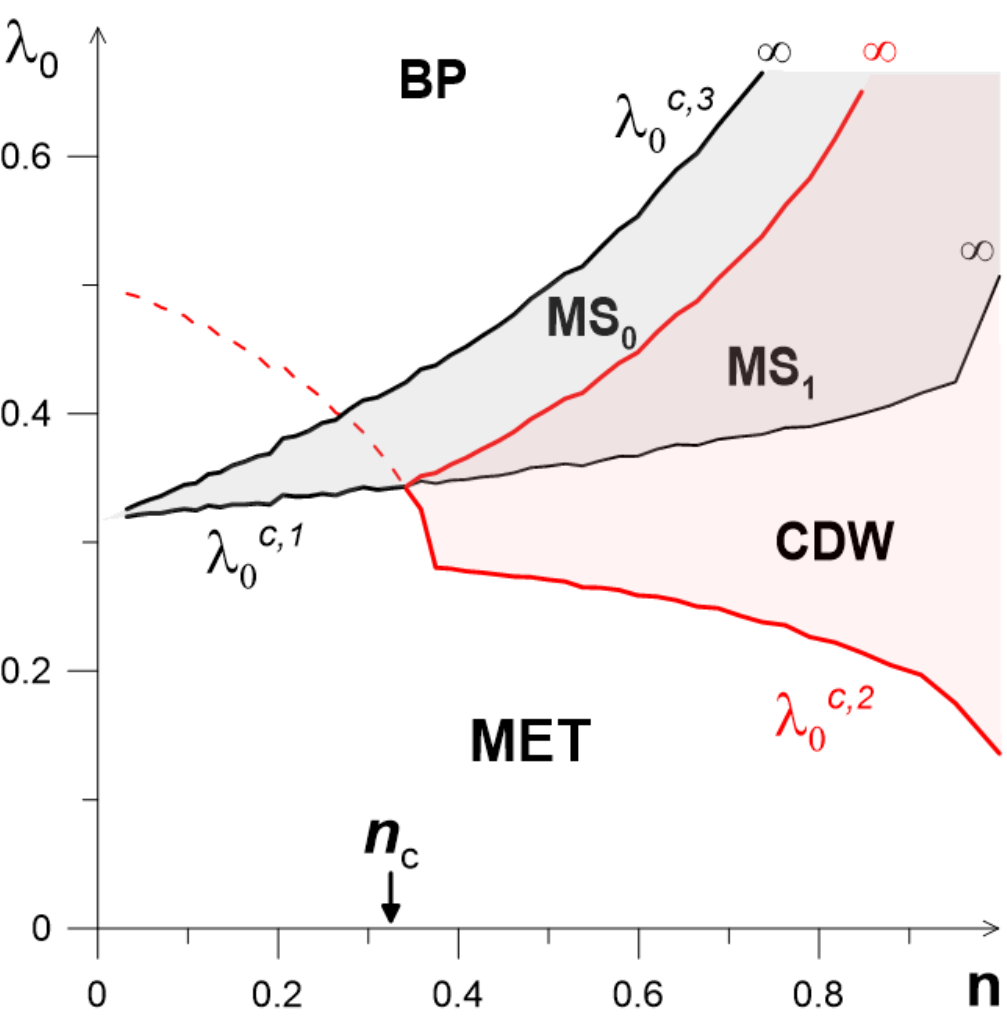}
\caption{The variational  phase diagram for  the 3D  (left) and 2D (right) e-ph system with
dispersionless Einstein phonon.
The analytical  phase diagram is identical, except  that fermions are treated as spinless and
the bipolaron state is replaced by a polaron one.
Phase diagram notations:
FL, a Fermi liquid, described by MET;
BP, a homogeneous bipolaron state;
CDW, a charge-ordered state of electrons; MS, mixed states, in which both the bipolaron and electronic  components are present. In MS$_0$ the electronic component is homogeneous and in MS$_1$  it is CDW-ordered
(shaded).
}
\label{fig:phase_diag_ms1}
\end{figure}

However, as argued by Dyson \cite{Dyson},
a perturbative expansion in the
coupling for a  continuous-space system in 3D is  prone to having a  zero radius of convergence.
In physical terms, the system may collapse into a state with an infinite density
because the potential energy gain $\propto n^2$  wins over the kinetic energy increase $\propto n^{5/3}$.
In 2D, kinetic and potential energies have the same $n^2$ scaling with density,
but  FL collapse may still occur.
For a lattice model, the Pauli principle limits electron density to two particles per site,
and the ``collapsed" state of spin-full fermions should be viewed as that of
bipolarons (BP)---local electron pairs that form a bound state with lattice
distortions (single polarons for spinless fermions).
In this situation, Dyson's
 considerations would lead to a finite radius of convergence, but a  collapse of the  FL state may still occur.
This  scenario of breaking MET is totally independent from the one
associated with the vanishing of the dressed phonon frequency.
\begin{figure}[htbp]
	\centering
    \noindent
	\includegraphics[width=0.4\linewidth]{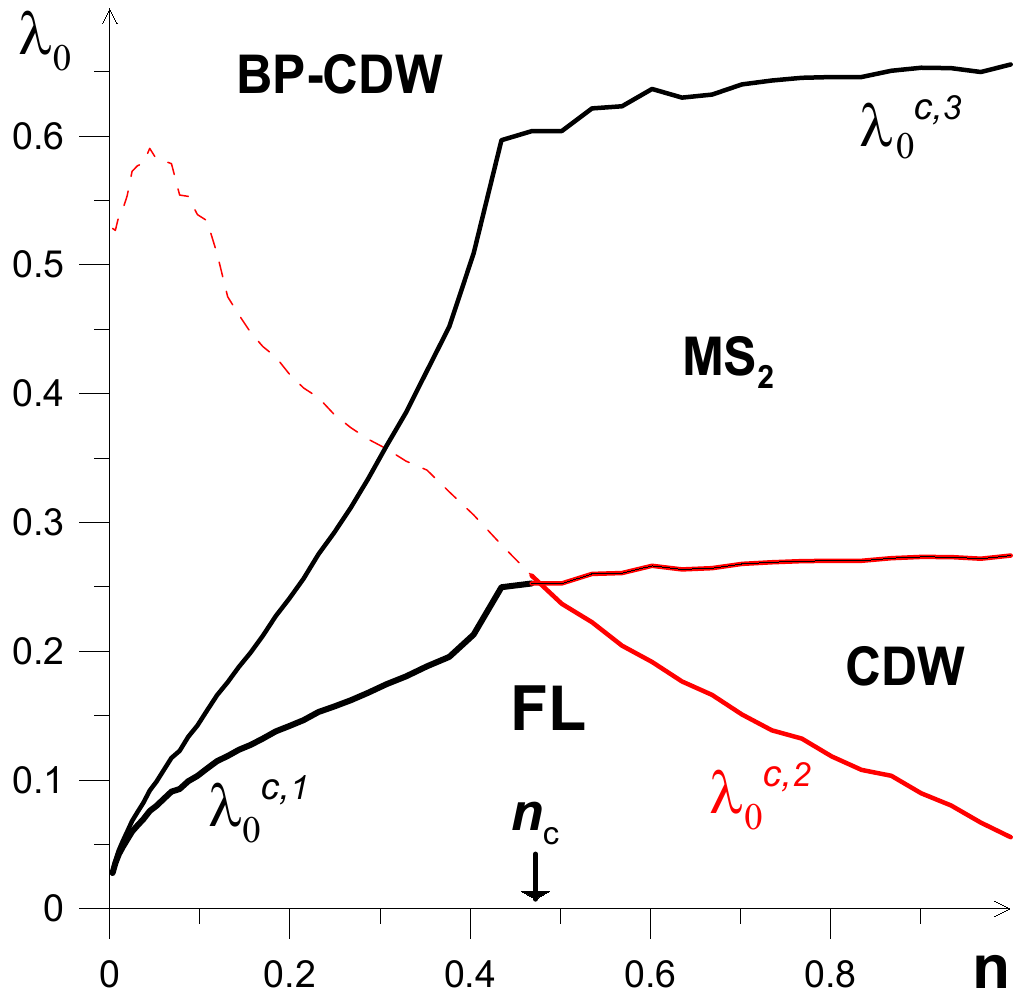}
	\includegraphics[width=0.4\linewidth]{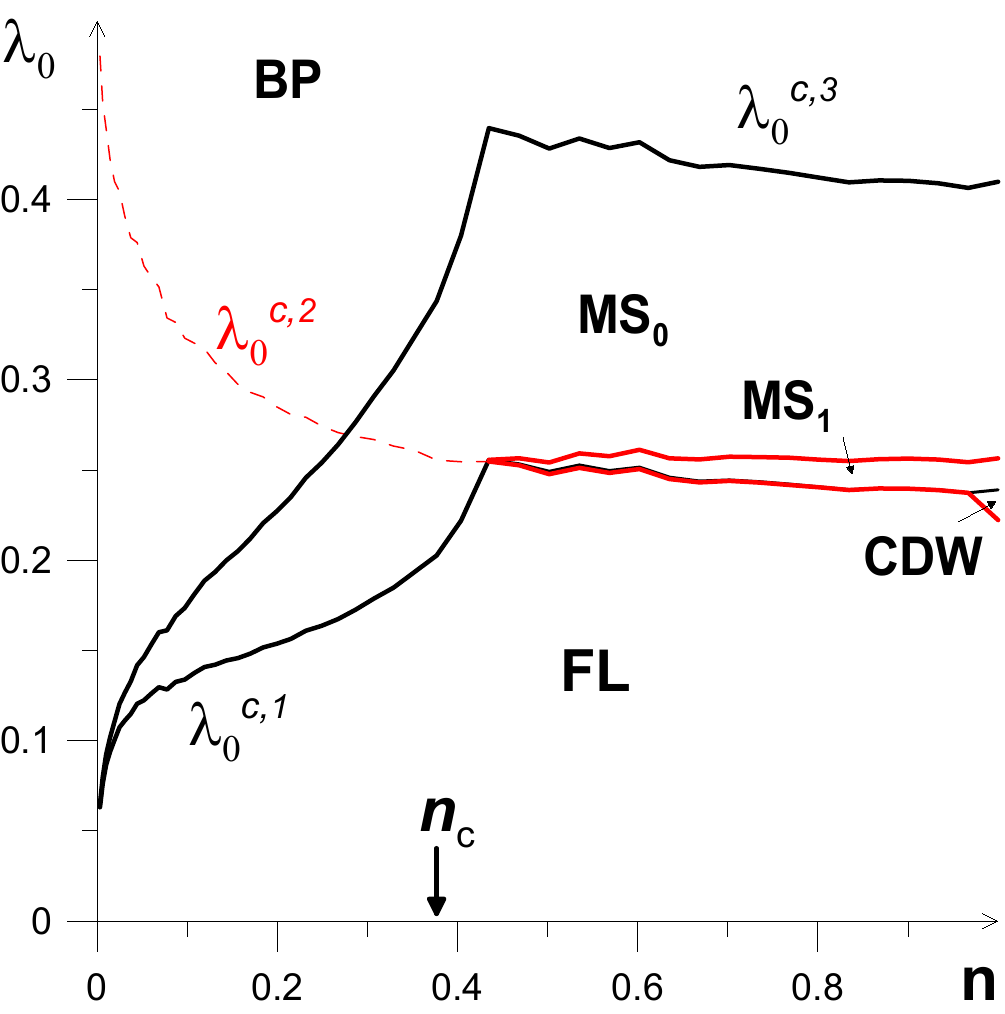}
\\
	\includegraphics[width=0.4\linewidth]{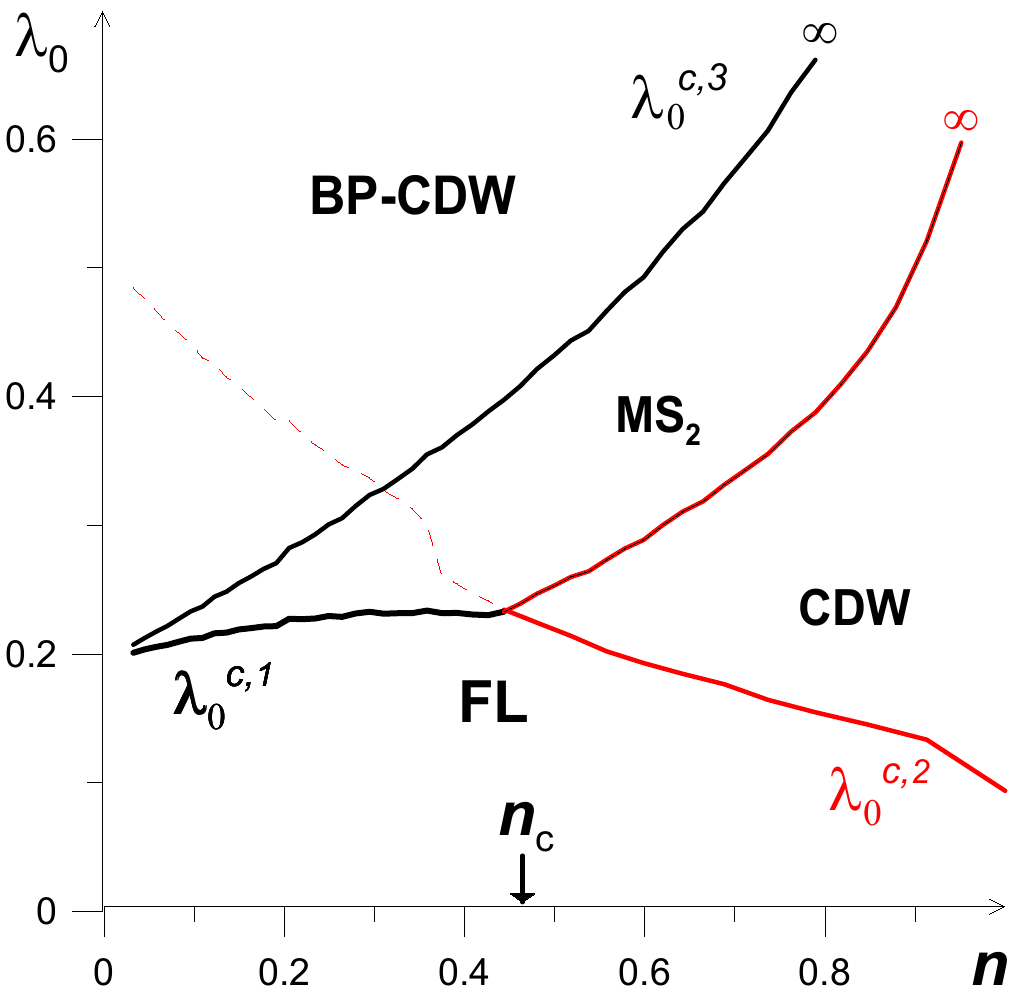}
	\includegraphics[width=0.4\linewidth]{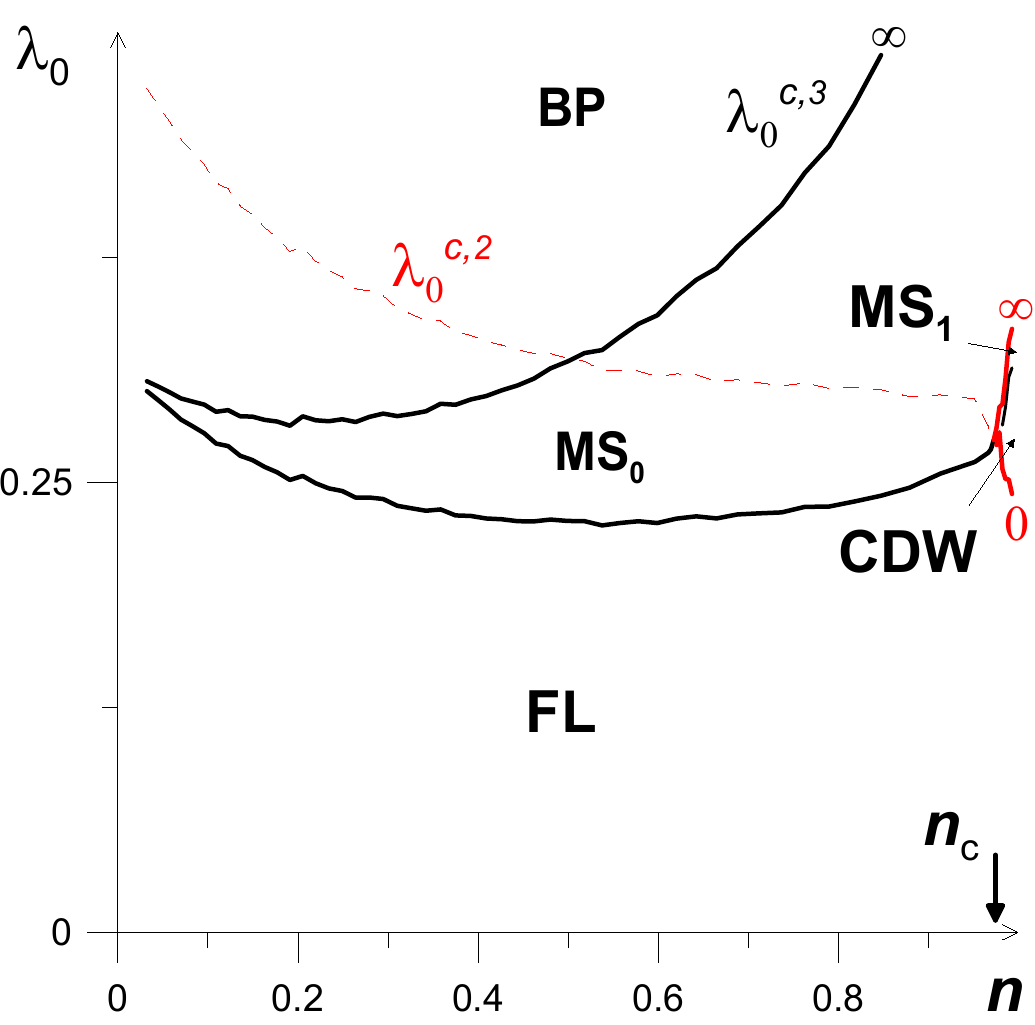}
\caption{
Numerical results for 3D (upper row)and 2D (lower row) fermions with tight-binding dispersion $\epsilon_{\mathbf k}= -2t \sum_{\alpha=1}^{D} \cos (k_{\alpha} a)$
coupled to phonons with dispersion  $\omega_0({\mathbf q}) = \omega_0 + 2 \tau \sum_{\alpha =1}^{D} [1- \cos (k_{\alpha}a)]$. We use units such that
$t=1$ and $a=1$, and set $\omega_0/t=0.1$. For $\tau = 0$ the spectrum is dispersionless, while for positive/negative $\tau$ its minimum is located at the origin/the zone boundary.
We set $\tau / t=-1/240$ in left panels, $\tau / t=1/120$ in right panels and  $\tau =0 $ in
Fig.~\ref{fig:phase_diag_ms1}.
For the left panel, we consider a checkerboard arrangement of bipolarons, labeled as BP-CDW.
For the right  panel and in Fig.~\ref{fig:phase_diag_ms1} we consider a homogeneous BP state.
The mixed states labeled $MS_i$  contain both FL and BP components, with equal chemical potentials but different volume fractions, which vary in opposite directions as $\lambda_0$ increases.
MS$_0$ is a mixture of the BP and FL states;
MS$_1$  of the BP and CDW states;
MS$_2$  of the BP-CDW and FL states.
At $\lambda_0 = \lambda_0^{c,1}(n)$,
the volume fraction of a BP/BP-CDW  vanishes,
at $\lambda_0 = \lambda_0^{c,3}(n)$, the volume fraction of a FL/CDW vanishes.
At $\lambda_0^{c,2}$ (red line) the phonon spectrum in the FL state
softens at some ${\mathbf q}_c$ and symmetry related momenta.
This line is  meaningful only when $\lambda_0^{c,2} < \lambda_0^{c,1}$.
When this holds, the system first develops a CDW order in between $\lambda_0^{c,2}$ and $\lambda_0^{c,1}$,  and the MS emerges out of a CDW state. In this situation,  $\lambda_0^{c,2}$
shows a re-entrant behavior  inside the mixed state.
}
\label{fig1}
\end{figure}

In this Letter,  we report results of variational and analytical studies
of a lattice e-ph system with
tight-binding electron dispersion of bandwidth $W$ and coupling (\ref{Hint}).
We show that, for  densities $n$ below a certain $n_c$,
a polaron/bipolaron state
\cite{Alexandrov,*Aleksandrov_2010, Mott,*Mott_1,Mahan00,polarons1998,polarons2000,Krezin2009}
emerges well before phonon softening.
At small $n$ or near-full band filling, when the
dispersion can be approximated as parabolic, this happens in 3D already at weak coupling,
as anticipated from  Dyson's argument.
We show that, for $0<n<2$ ($0<n<1$ for spinless fermions), FL remains stable up to
a finite $\lambda^{c,1}_0$ and at larger $\lambda_0$ is replaced by a mixed state, in
which some fermions still display FL behavior,
but their density $n_{FL}$ gradually decreases from $n$, while the rest of  the fermions, with density
$n_{bp}$,  form localized polarons/bipolarons. In this state, the density of states (DOS), $N(\omega)$,
displays  pseudogap behavior, the Green's function has both poles and zeros,
and the canonical Luttinger theorem is broken.
The density $n_{FL}$ vanishes at some larger $\lambda^{c,3}_0$,
and at even larger $\lambda_0$ the system enters a pure polaron/bipolaron state, in which all
electrons form bound states with lattice distortions.
The DOS in this state consists of two continua, centered at positive and negative frequency, each with width $W$,  and narrow patches of DOS of heavy polarons  at smaller $|\omega|$,  down to $\omega =0$.
For spinless fermions, narrow patches extend down to $\omega =0$;  for spin-full fermions,
$N(\omega)$ vanishes below a certain $\omega$.
We also find that closer to half filling,  the leading instability of the FL upon increasing coupling is
phonon softening, leading to CDW electronic order,  and a polaron state emerges out of the CDW.
For these dopings, there exists additional line in the mixed state above which the
CDW order vanishes (see Figs. \ref{fig:phase_diag_ms1} and \ref{fig1})
\footnote{To avoid misunderstanding, the ground state  for spin-full fermions is a superconductor.
We assume that superconducting $T_c$ is small and our analysis holds at $T>T_c$, which is
low enough to avoid other temperature effects.}. 
Below we present some details of our variational and analytical calculations.  We did variational calculations for
both dispersing and flat $\omega_0 (q)$ and analytical calculations for a flat $\omega_0 (q) = \omega_0$.
For the latter case, we obtained identical variational and analytical phase diagrams, 
which we show in  Fig. \ref{fig:phase_diag_ms1}. We caution that beyond our computational approach, 
 a checkerboard polaron state  likely has a lower energy near $n=1$ than a homogeneous bipolaron state, even for a flat Einstein phonon.  Full details of our calculations are presented in the companion paper~\cite{companion}.

{\it {Variational analysis}}~~~In Figs.~\ref{fig:phase_diag_ms1}
and \ref{fig1},
we present results of the variational analysis for fermions on the simple cubic (3D)
and square (2D) lattices with tight-binding
dispersion. We show representative cases for
three types of $\omega_0 (q)$: the dispersionless one
(Fig.~\ref{fig:phase_diag_ms1}),
the one with a minimum at zero momentum,  and the one with a minimum at
 ${\mathbf Q}=(\pi,\pi,\pi)$ in 3D and $(\pi,\pi)$ in 2D (Fig.~\ref{fig1}). 
To obtain the phase diagram, we checked at which $\lambda_0$
(i) the chemical potential of a BP state, $\mu_{BP}$  becomes the same as the chemical potential of a FL state, $\mu_{FL}$,
at vanishing FL density (the upper end of the mixed state),
(ii) $\mu_{FL}$ at the full density $n$  becomes the same or higher than $\mu_{BP}$,
 provided the BP state has lower energy than the FL at the density of the BP state, $E_{BP} (n_{bp} ) \le E_{FL} (n_{bp} )$
(the  lower end of the mixed state),
and (iii) the dressed phonon spectrum softens to zero at some momentum.

The dominant contributions to the FL ground state energy
are the kinetic energy of noninteracting itinerant fermions,
$E_{\text{kin}} = 2\sum_{\epsilon_{{\mathbf k}}<E_F} \epsilon_{{\mathbf k}}$,
and the Hartree potential energy,
$U = -N[g/\omega_0(0)]^2 n_{FL}^2/2$; i.e. $E_{FL}=E_{\text{kin}}+U$.
Other contributions to the  energy are small in the adiabatic parameter $\gamma$.
The chemical potential of the FL is $\mu_{FL}=E_F-[g/\omega_0(0)]^2 n_{FL}$.
A competing variational BP state is a set of
{\it {localized}} bipolarons with a density profile $n_i$,
whose Fourier transform is $n_q$.
Its exact energy $E_{BP}$ is obtained from the shifts of harmonic modes
\begin{equation}
E_{BP} = -\frac{g^2}{2} \sum_{{\mathbf q} } \frac{|n_{{\mathbf q}}|^2}{\omega^2_0({\mathbf q})} .
\label{Ebi}
\end{equation}
We consider two BP states. One, abbreviated BP in the figures,
is a homogeneous state with the highest possible density
$n_{bp}=2$.  Its ground state energy is $E_{BP}^{(a)} = - N_e g^2/\omega_0^2(0)$ and the chemical potential is
$\mu_{BP} = E_{BP}^{(a)}/N_e$.
Another state, abbreviated as BP-CDW, is a $n_{bp}=1$ state with the checkerboard arrangement
of sites with $n_i=0$ and $n_i =2$.
 Its energy is $E_{BP}^{(b)} = -N_e (g^2/2) [1/\omega_0^2(0) + 1/\omega_0^2(\mathbf{Q})]$
and the chemical potential is $\mu_{BP} = E_{BP}^{(b)}/N_e$.
The BP state has lower energy when the minimum of the bare phonon dispersion is
at zero momentum and the BP-CDW state wins when the minimum of $\omega_0 (\mathbf{q})$ is at $\mathbf{Q}$.
For a dispersionless  Einstein phonon, any configuration of localized pairs has the same energy
$E_{BP} = - N_e g^2/\omega_0^2$ and chemical potential $\mu_{BP} = -g^2/\omega_0^2$.
We didn't include terms which remove this degeneracy (see Refs.~
\cite{companion,Chakraverty1,Chakraverty2}) and
lower the energy of the  true ground state. In this respect, our variational analysis
establishes the upper bound on the coupling
at which the FL state  becomes unstable toward bipolarons.
On general grounds, we expect the BP state
at small densities and  the BP-CDW state near half filling.

There are two key critical lines in each
phase diagram in Figs.~\ref{fig:phase_diag_ms1} and \ref{fig1}:
at $\lambda_0^{c,1}$, $\mu_{FL}(n) \ge \mu_{BP} $
and the mixed  (likely phase separated) state FL/BP or FL/BP-CDW  emerges;
at $\lambda = \lambda_0^{c,2}$ the phonon spectrum softens at some momentum, and the CDW state emerges.
Each is a true instability line when the corresponding critical value of $\lambda_0$ is smaller.
Namely, when $\lambda_0^{c,1} < \lambda_0^{c,2}$, the FL
becomes unstable toward bipolarons despite MET remaining internally stable;
when $\lambda_0^{c,2} < \lambda_0^{c,1}$, the system first develops CDW order.
In 3D, we find that at small densities,
when the Fermi surface is nearly spherical, $\lambda_0^{c,1}$  tends to zero as  $n^{1/3}$
due to the vanishing $N_F \sim p_F \sim n^{1/3}$ while $\lambda_0^{c,2}$  remains finite.
As a result, at $n \to 0$,  the FL  becomes unstable against bipolarons already at
infinitesimally small $\lambda_0$, implying that  the continuum  limit
 of a 3D  e-ph system is not described by MET at any $\lambda_0$.
In 2D, $\lambda_0^{c,1}$  tends to a finite value
at $n \to 0$ (see Fig. \ref{fig1}; $\lambda_0^{c,1} = 1/\pi$ for a dispersionless phonon),
because in this limit $N_F$ does not depend on the Fermi momentum.
Still, this value is smaller than $\lambda_0^{c,2} \approx 0.5$;
i.e., at small $n$, the FL again becomes unstable against  bipolarons before the phonon spectrum softens.

We find that when $\lambda_0$ increases above
$\lambda_0^{c,1}$,  the
density of the FL component $n_{FL}$ decreases and some  fermions move into
the BP (or BP-CDW)  state.

The chemical potentials $\mu_{BP}$ and $\mu_{FL}$ remain equal within the mixed phase,
as required by the Maxwell construction. The mixed phase
ends at  $\lambda_0 = \lambda_0^{c,3}$, at which
$n_{FL}=0$. At larger $\lambda_0$, the  entire system  consists of BP or CDW-BP.
 In 2D, $\lambda_0^{c,3}$ diverges at half filling because the DOS for the tight-binding dispersion
in 2D diverges at $n=1$. For densities $n$ at which
 $\lambda_0^{c,2} < \lambda_0^{c,1}$, CDW order develops
first at $\lambda_0=\lambda_0^{c,2}$ and the mixed state develops out of the CDW state.
In this range,  the line
$\lambda_0^{c,2}$  shows a re-entrant behavior
inside the mixed phase (upper solid red line in Fig. \ref{fig1})
and determines where the CDW order disappears.
Note that in the lower right panel in Fig.  \ref{fig1}, $\lambda_0^{c,2} < \lambda_0^{c,1}$ only very near half filling,
and  in 3D the mixed state exists only in a very narrow range of couplings.
\begin{figure}[t]
	\includegraphics[width=0.9\linewidth]{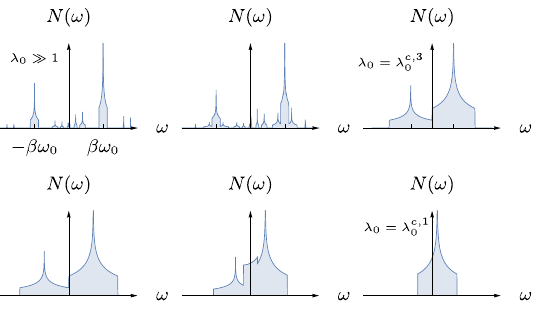}
    \caption{
DOS of spinless fermions in a 2D system with dispersionless phonon
for a generic density $n<1/2$.
The width of the peak at $\omega = \beta \omega_0$ is set by $W$,
the width of the peaks near $\omega = 0$ is exponentially small $\sim We^{-\beta}$.
 The peak centered at $\omega =0$ consists of two independent parts
 at $\omega >0$ and $\omega <0$,  for each $N(\omega =0) =0$.
As $\lambda_0$ decreases, polaron peaks are absorbed by the continua one by one.
For $n=0$ ($n=1$),  $N(\omega)$ is nonzero only  for  $\omega >0$ ($\omega <0$).
Bottom: DOS evolution in the mixed phase, for   $\lambda_0^{c,1} \leq\lambda_0  \leq W N_F/2$.
The DOS displays  pseudogap behavior: two peaks at finite $\omega$ and a nonzero DOS in between.
 }
\label{fig:extra}
\end{figure}

{\it {Analytical results}}~~~
We consider spinless fermions with tight-binding dispersion in 2D and
dispersionless phonon with frequency $\omega_0$.
We set $\beta = g^2/(2\omega^3_0)$ to be large and vary
 $\lambda_0 = \beta \omega_0 N_F = 4\pi^2  (\beta \omega_0/W) K (1 - 4 \mu^2_{FL} (n)/W^2)$,
where $K(...)$ is the elliptic integral, by varying the
bandwidth $W$.
We adopt diagrammatic treatment departing from free fermions at arbitrary $\lambda_0$,  with the goal to describe the transformation from the FL to polaron state via an intermediate mixed phase. We obtain two sets of analytical results: one for $n =0$ and $n=1$,
and another for arbitrary $n$.  We show the details in the End Matter and present here the summary.
For $n=0$ and $n=1$,  we demonstrate that the exact DOS of the Holstein model in the atomic limit
$W=0$ ( $\lambda_0 = \infty$)
\cite{lang_firsov,Holstein1959,Mahan00}
is reproduced by the eikonal approach, which treats self-energy and vertex corrections on equal footing.
The eikonal series (continued fractions) strongly evolves at an order
of perturbation theory 
$m \approx \beta $,
which we associate with the number of phonons in the polaron state at the smallest $\omega$.
We then extend  the diagrammatic analysis to finite $W$  and show that DOS can be approximated by a
set of exponentially ($e^{-\beta}$) narrow patches  at low energies, $\omega = l\omega_0$  which describe heavy polarons
($l =0, 1, 2, \dots$  for $n=0$,   $l =0, -1, -2, \dots$ for $n=1$ and $|l| \ll \beta$),
and  a nearly free-particle continuum of width $W$, centered at much larger $ |\omega| = \beta \omega_0$ (Fig. \ref{fig:extra}, top panel,
$N(\omega) \neq 0$ at $\omega >0$ for $n=0$ and at $\omega <0$ for $n=1$).
As $W$ increases, the lower end of the continuum moves toward $\omega =0$, absorbing  low-energy polarons  one by one.
A similar behavior has been reported in DMFT studies~\cite{Ciuchi_1997}.
The last polaron at $\omega =0$ is absorbed at $\lambda_0^{c,1}(n=0) =1/\pi$  and, at smaller $\lambda_0$,
the system is in the FL state described by MET.
   \begin{figure}[]
	 \includegraphics[width=0.9\linewidth]{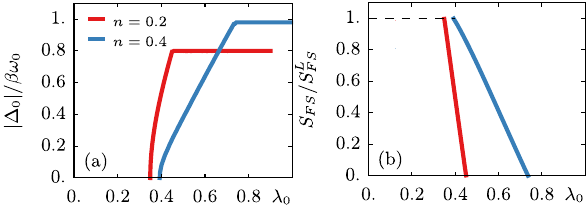}
	 \caption{(a) The condensate order parameter $|\Delta_0|$ vs $\lambda_0$.  For $\lambda_0 > \lambda_0^{c,3} =  (W N_F/2)$,
$|\Delta_0|$ is independent on $\lambda_0$. At smaller $\lambda_0$,
 it decreases and eventually vanishes at $\lambda_0 = \lambda_0^{c,1}$ (see text). (b): the area of the Fermi surface $S_{FS}$ in the mixed state, normalized by the Luttinger area in the FL state $S_{FS}^L$. }
	 \label{fig:delta_vs_lambda}
	 \end{figure}
At finite $n$, we argue that one has to introduce an ancilla fermion ${\tilde c}$
with the average density $1-n$, in addition to the physical $c$ fermion with the average density $n$, 
to account for the fact that there are two types of excitations at a given site
because it can be either occupied or empty.
We argue that the ground state energy can be lowered by introducing  a condensate  order  parameter
$\Delta=\langle c^{\dagger}c\rangle$, whose value we obtain by minimizing the ground state energy as a function of $\Delta$.
Diagonalizing the effective quadratic Hamiltonian and re-expressing the
e-ph interaction in Eq. (\ref{Hint}) in terms of new fermions,
we exactly split the full Hamiltonian into two parts: one is the same as at $n=0$ and the other as at $n=1$.
Eikonal calculation at $W=0$ for each part reproduces the exact DOS for the physical $c$ fermion~\cite{lang_firsov,Holstein1959,Mahan00,Berciu_2006}, consisting of two sets of $\delta$ functions at $\omega \geq 0$ and $\omega \leq 0$,
the first with the overall factor $1-n$ and the second with the overall factor $n$.
We then extend the calculations to  finite $W$. Now the choice of
condensate momentum $\mathbf{q}$ matters.

Similar to the variational analysis, we consider a homogeneous polaron state with $q=0$ and the
checkerboard state with $\mathbf{q}=\mathbf{Q} = (\pi,\pi)$. Here we
discuss the homogeneous state with  $\Delta_{q=0} = \Delta_0$;
the  case $\Delta (\mathbf{q}=\mathbf{Q})$ is analyzed in ~\cite{companion}.
For $q=0$, we find again that the diagonalized Hamiltonian exactly splits into two parts: one is the same as at $n=0$, and the other at $n=1$.
This implies that polarons remain  fully localized in the sense that there are no density fluctuations between occupied and non-occupied states.
We find that the DOS at large but finite $\lambda_0$ again can be approximated by a
set of exponentially narrow patches of heavy fermions at $\omega = l \omega_0$ ($l =0, \pm1, \pm 2, \dots$,   $|l| \ll \beta$),
and  two free continua at higher energies, centered at
$ \omega = \pm \beta \omega_0$, each of width $W$ (Fig. \ref{fig:extra}, top panel.)
The DOS  vanishes strictly at  $\omega =0$ (in the middle of the patch $l=0$) implying that there is no Fermi surface in the polaron state.
As $W$ increases and $\lambda_0$ decreases,  the lower edges of the two continua extend to smaller
$|\omega|$ and absorb low-energy polarons one by one.
As long as the continua are separated,
$\Delta_0 = \beta \omega_0 [4n(1-n)]^{1/2}$ is independent of $\lambda_0$
(Fig. \ref{fig:delta_vs_lambda} a).  The  edges of the two continua merge at $\omega =0$  at $\beta \omega_0 = W/2$, i.e., at
$\lambda_0 = \lambda_0^{c,3} = W N_F/2$.   At smaller
$\lambda_0$, the DOS displays pseudogap behavior (Fig. \ref{fig:extra}, bottom panel.)
A similar behavior has been detected in DMFT studies~\cite{Millis_1996,*Millis_1996_a}.
We find that in this range of $\lambda_0$,   the ground state can be viewed as a mixed state, in which a portion
of the system with density $n_p$ is in the homogeneous polaron state and the other portion, with density $n-n_p$ is in a FL state.
  The values of  $n_{FL} (\lambda_0, n)$  are obtained by minimizing the sum of the polaron and FL energies.
  At a given $n$, $n-n_{FL}$ decreases with decreasing $\lambda_0$ and vanishes along with $\Delta_0$
  at $\lambda_0 = \lambda_0^{c,1} (n) < W N_F/2$.
At smaller $\lambda_0$, the ground state is a FL described by MET.  We obtain the Green's function in the mixed
state and show that it has both poles and zeros. The FL fermions in the mixed state have a Fermi surface, whose
area is given by  $n_{FL}<n$
[Fig. (\ref{fig:delta_vs_lambda}b)].
We view this result as an indication that the canonical Luttinger theorem, by which the area must be  equal to $n$,
is violated in the mixed state. This result is  expected,  as the U(1) condensate order parameter $\Delta_0$ is nonzero in the mixed state.
We argue  that  the transition from the FL to the mixed state at $\lambda_0^{c,1} (n)$ can be viewed as a $q=0$ charge instability in a FL,
in which translation symmetry gets broken as
as a fraction of  fermions get localized by  forming polaron bound states that do not participate in the Luttinger count.
There is a certain similarity between our mixed state and spin-liquid states like
FL$^*$, suggested for hole-doped cuprates (see e,g., \cite{Bonetti_2025,Senthil_2003,Sachdev_2010}
\footnote{ We caution that in our analysis we didn't include fluctuations of the U(1) order $\Delta_0$.  If fluctuations destroy long-range order in $\Delta_0$ by producing a finite correlation length $\xi$, the Luttinger theorem is not broken, yet above a characteristic frequency $\omega* \propto \xi^{-1}$, the
  spectral function $A(k, \omega)$   develops strong peaks at momentum  $k$ which correspond to the small Fermi surface in our mean-field analysis~\cite{Chubukov1995,*Vilk1997_a}.  }  ) .
Similarly to the variational analysis, for $n$ near half filling we find that the first instability upon increasing $\lambda_0$ is into the CDW state and the $q=0$  polaron state emerges at larger $\lambda_0$ from the CDW state. The analytical phase diagram  is identical to the one in Fig. \ref{fig:phase_diag_ms1} after an adjustment from bipolarons to polarons.

{\it {Conclusions}}~~~ In this Letter, we considered a system of spin-full and spinless fermions, interacting with an optical phonon in the adiabatic limit.  For weak/moderate coupling, such  a system is  believed to be in a FL regime described by MET.
We show that at small  electron  density,  the  FL state in 3D is unstable toward bipolarons/polarons already  at small coupling $\lambda_0\ll 1$, when MET is stable against low-energy fluctuations.  In 2D,
the FL state  becomes unstable toward bipolarons/polarons  at a finite $\lambda_0$, but still when MET is internally stable.  We showed that the transformation from a FL to a bipolaron/polaron state occurs via an intermediate mixed phase, in which a FL still exists, but with a smaller density/smaller Fermi surface, while the remaining fermions form
localized  heavy bipolarons/polarons.
At larger densities, the instability toward CDW may come first, and  bipolarons/polarons form out of a CDW state.
Overall, our results imply  that  the emergence of bipolarons/polarons imposes the most severe limitation
on the range of applicability of MET for almost all fermionic densities.

{\it {Acknowledgement. }}~~~We acknowledge with thanks useful discussions with 
 B. Altshuler, M. Berciu, E. Berg, S. Ciuchi, M. Fabrizio, R. Fernandes, S. Fratini, S. Ilani,  
T. Heikkila, W. Metzner, C. Murthy, H-Y Kee, M. Kiselev,  S. Kivelson, A. Millis, N. Nagaosa, P. Nosov, 
R. Ojajaervi, Ph. Phillips, M. Randeria,  S. Sachdev,  M.V. Sadovskii,  G. Sangiovanni, D. Senechal,  J. Schmalian, A. Stern,  B. Svistunov, A-M Tremblay, Y. Wang,  M. Ye, Y. Yu,  and S.-S. Zhang.
 A. C. was supported by the U.S.\ Department of Energy, Office of Science, Basic Energy Sciences, under Award No.~DE-SC0014402.
 A. A. was supported by the DOE-Office Of Science, DE-SC0026038.
A. C. and N. P. acknowledge support from the Simons Foundation grant SFI-MPS-NFS-00006741-07 for the Simons Collaboration on New Frontiers in Superconductivity.
A.C. and N. P. thank the Aspen Center for physics for the hospitality at the early stages of this project.  Aspen Center for Physics  is supported by National Science Foundation grant PHY-2210452.

\bibliography{ref_phonons_short}

\pagebreak
\section{End Matter}

In this End Matter we present some details of our analytical treatment of a 2D systems of spinless fermions coupled by
Eq. (\ref{Hint}) to a phonon with a dispersionless $\omega_0$.   More details are presented in Ref. \cite{companion}.
We use diagrammatic treatment as our goal is to describe the transformation from a FL at small $\lambda_0$ to a polaron state at large $\lambda_0$.
Our goal is twofold. One is to demonstrate that a polaron instability comes from high orders of perturbation  theory, when characteristic energy for multiple phonon scattering is of order bandwidth.  Such an instability  cannot be detected  within a  conventional low-energy perturbation theory as it  only includes contributions from low-energy fermions located well within the bandwidth.  Another is to show details of our description of  the violation of the famous Luttinger theorem in systems without CDW or SDW order.

At zero filling $n=0$ and at full filling $n=1$  we
 depart from free fermions and sum up infinite series of
  self-energy and vertex corrections taken on equal footings (the approach is commonly known as eikonal~\cite{Levy1969,Lee1973,Efros1971,Sadovskii1974,*Sadovskii1974SS,*Sadovskii_review,*Sadovskii_extra,*Sadovskii_extra,
  Vilk1997,Schmalian1998,*Schmalian1999,Tchernyshyov1999,Rohe_2005,Sedrakyan2010,Yamase_2012,*Yamase_2016,Ye2023,*Ye2023_1,
  Posazhennikova2003,Kiselev2009,*Efremov2022}).
In the atomic limit, $W=0$, we find
\bea
 &&G^{n=0}(\omega) = \frac{1}{\omega + i\delta} {_1}F{_1} (1, 1 - \frac{\omega + i\delta}{\omega_0}, - \beta), \nonumber \\
&&G^{n=1}(\omega) = \frac{1}{\omega + i\delta} {_1}F{_1} (1, 1 + \frac{\omega + i\delta}{\omega_0}, - \beta),
  \label{k_4}
     \eea
where  ${_1}F{_1} (a,b,c)$ is the Kummer confluent hypergeometric function, $\omega_0$ is a phonon frequency, and $\beta = g^2/(2\omega^3_0)$, where
 $g$ has been defined in (\ref{Hint}) (see also
 \cite{SadovskiiBook,Kuchinskiy_2024}).
The Kummer function
 has an infinite set of poles (polaron $\delta-$functional peaks)  at $\omega = l \omega_0$ with $l = 0,1,2..$ for $n=0$ and $l = 0,-1,-2..$ for $n=1$.  Eq. (\ref{k_4}) reproduces the exact result for the fermionic DOS in the atomic limit\cite{lang_firsov,Holstein1959,Mahan00}.
  The poles at small $\omega \sim \omega_0$ have exponentially small residue $Z_m \sim e^{-\beta}$.
   By analyzing the diagrammatic series, we explicitly verified that a polaron at $\omega =0$
    comes from high  order  $O(\beta)$  terms in the eikonal expansion and is a bound state of an electron and  $O(\beta)$
    phonons (to see that the order $O(\beta)$ is  special, it is convenient to re-express diagrammatic series as continuous fractions~\cite{Ciuchi_1997,Goodvin2006,Berciu_2006}).
   The  poles at
   $\omega \approx \pm \beta \omega_0$, i.e., at $l \approx \pm \beta$  have larger
   $Z_m \sim 1/\sqrt{\beta}$.

    We then add a fermionic dispersion set by $W$.  We find that it affects  polaron peaks at small $\omega$ and  at $\omega \approx \beta \omega_0$ differently.   For the former, the dispersion converts
    $\delta-$functional peaks into exponentially narrow patches of width $W e^{-\beta}$,  between which the DOS vanishes.
    This agrees with the earlier results~\cite{Ranninger_1992,*Ranninger_1993,Kornilovitch_2002,Berciu_2006,esterlis_18,*esterlis_19, *Nosarzewski2021,*murthy_2023}.
      For the latter, individual patches start overlapping at $W \sim \sqrt{\beta} \omega_0$ (Ref. \cite{Berciu_2006}) and at larger $W$
       form a
       single patch with width $W$. The total residue of this patch is $Z =1 - O(e^{-\beta})$,
        i.e., to  high accuracy it can be viewed as a DOS of a free dispersing fermion.
       The picture of DOS consisting of a free-fermion peak with width $W$, centered at $\pm \beta \omega_0$ ($+$ for $n=0$ and $-$ for $n=1$) and exponentially narrow patches of polarons at smaller $\omega$ is valid for $\sqrt{\beta} \omega_0 \ll W \ll \beta \omega_0$   As $W$ increases toward $\beta \omega_0$, the free-fermion DOS gets broader,  its lower end stretches toward $\omega=0$ and absorbs patches of heavy fermions one by one (Fig. \ref{fig:extra} top).
       A similar behavior has been found in DMFT studies
       \cite{freericks1993, Millis_1996,*Millis_1996_a, Ciuchi_1997, Ciuchi_1993,*Ciuchi_1998,*Ciuchi_2003,*Ciuchi_2006,*fratini2021,*Ciuchi_2025}.
       The last patch at $\omega =0$ is absorbed at $\beta \omega_0 = W/2$ ($\lambda_0  = 1/\pi$).
         This is the end point of the polaron state.  At larger $W$,  the system is in the FL phase and the DOS $N(\omega)$ is continuous between $\omega =0$ and $\omega = \pm W$.

At a finite $n$, the analysis is more involved because
  already
  in the atomic limit, each given site  can be in one of two states:  occupied or empty.
   Accordingly, there is a  separate set  of excitations above each of the two states, each has to be  described by a separate fermion.
  This agrees with the form of the  exact DOS, which at  $0<n<1$ consists of two sets of $\delta-$functional peaks, one at positive $\omega$ and another at negative $\omega$ (Refs.\cite{lang_firsov,Holstein1959,Mahan00}).
 The set at $\omega >0$ is the same as at $n=0$, but  the residue has an additional factor $1-n$; the set at $\omega <0$ is the same as at $n=1$, but with   residue $n$. The DOS has two maxima at $\omega = \pm \beta \omega_0$.  To   satisfy the   requirement that there must be two types of excitations,
  we introduce a physical fermion $c$ describing quasiparticles with  density $n$ and an
   ancilla  fermion ${\tilde c}$ describing quasiparticles with density $1-n$.  The chemical potential of a physical fermion, corrected by the Hartree term, is
   $\mu_n = -\beta \omega_0 (1-2n)$,  the one for the ancilla fermion is ${\tilde \mu}_n = -\beta \omega_0 (1-2(1-n)) = - \mu_n$.
   There is a similarity between our approach and bimodal distribution of phonon displacements, found in~\cite{Millis_1996,*Millis_1996_a}.
   The physical and ancilla fermions both couple to phonons as in Eq. (\ref{Hint}). The effective Hamiltonian is
 \beq
   H' =  -{\mu_n} \left(c^\dagger c - {\tilde c}^\dagger {\tilde c} \right)  + \frac{g}{\sqrt{2\omega_0}} \left(c^\dagger c + {\tilde c}^\dagger {\tilde c}\right) (b + b^\dagger).
   \label{10_d}
   \eeq
   To eliminate phonon-mediated coupling between the physical and ancilla fermion, we  simultaneously
  have to introduce the additional counterterm
   $H_{c,{\tilde c}} =  2 \beta \omega_0 c^\dagger  c {\tilde c}^\dagger {\tilde c} $.
  The full Hamiltonian is $H = H' + H_{c,{\tilde c}}$
We assume and then verify  that we can lower the  ground state energy by
  introducing a $U(1)$.
order parameter  (a condensate)  $ \Delta = - 2 \beta \omega_0  \langle c^\dagger {\tilde c} \rangle$.
 Using  this term to decouple $H_{c,{\tilde c}}$, we find
\bea
   H'  &=& \frac{|\Delta|^2}{2\beta \omega_0} -{\mu_n} \left(c^\dagger c - {\tilde c}^\dagger {\tilde c} \right) + \Delta c^\dagger {\tilde c} + \Delta^* {\tilde c}^\dagger c \nonumber \\
   && + \frac{g}{\sqrt{2\omega_0}} \left(c^\dagger c + {\tilde c}^\dagger {\tilde c}\right) (b + b^\dagger).
   \label{10}
   \eea
Diagonalizing the quadratic part of $H'$ by Bogolyubov rotation to new fermions $\alpha$ and ${\tilde \alpha}$, we obtain
\bea
  && H^{'} =  \frac{|\Delta|^2}{2\beta \omega_0} + E \left( \alpha^\dagger \alpha - {\tilde \alpha}^\dagger {\tilde \alpha} \right) \nonumber \\
&& + \frac{g}{\sqrt{2\omega_0}} \left(\alpha^\dagger \alpha + {\tilde \alpha}^\dagger {\tilde \alpha} \right) (b + b^\dagger) = H^{'}_{\alpha} + H^{'}_{{\tilde \alpha}},  \label{11}
   \eea
   where $E =({\mu_n}^2 + |\Delta|^2)^{1/2}  = ((\beta \omega_0 (1-2n))^2 + |\Delta|^2)^{1/2}$.
     The full Hamilonian also contains $H_{c,{\tilde c}}$, re-expressed via $\alpha$ and ${\tilde \alpha}$.  This counterterm still eliminates the phonon-induced interaction between $\alpha$ and ${\tilde \alpha}$ fermions,   hence $ H^{'}_{\alpha} $ and $ H^{'}_{{\tilde \alpha}},$  in Eq. (\ref{11}) are truly decoupled.
The self-consistency condition on
 $\Delta$ yields
 $|\Delta| = 2 \beta \omega_0 (n(1-n))^{1/2}$. Substituting into $E$, we  $E = \beta \omega_0$,  independent on $n$.
We see that
 $H^{'}_{\alpha}$  is the same as $H'$ at $n=0$ and $H^{'}_{{\tilde \alpha}}$
 is the same as at  $n=1$. The two terms
 describe  realizations in which a given lattice site is occupied or empty.
   The corresponding fermions  $\alpha$ and ${\tilde \alpha}$ are linear combinations of the original and ancilla fermions.
The Green's function of the physical fermion $c$,   $G^{c} (\omega)$, is
 \beq
   G^{c} (\omega) =  (1-n) G^{\alpha} (\omega) + n G^{{\tilde \alpha}} (\omega).
  \label{qq_7}
   \eeq
 where  $G^{\alpha} (\omega) = G^{n=0} (\omega)$ and  $ G^{{\tilde \alpha}} (\omega)= G^{n=1} (\omega)$ are given by  Eq.  (\ref{k_4}).
 This  reproduces the exact Green's function in the atomic limit (see e.g. \cite{Mahan00}).

 We now add the  dispersion $\epsilon_k$ to physical and ancilla fermions $c_k$ and ${\tilde c}_k$.   Like in the main text, we assume here that the condensate  is homogeneous $\Delta_{q=0} =
  \Delta_0 =-2\beta \omega_0 (1/N) \sum_k \langle c^\dagger_{\mathbf k} {\tilde c}_{{\mathbf k}}\rangle$.  
   In terms of new operators $\alpha_k$ and ${\tilde \alpha}_k$ we now obtain
       \bea
   H' &=& \sum_{\mathbf k} \left[\left(\epsilon_{\mathbf k} + E\right) \alpha^\dagger_{\mathbf k} \alpha_{\mathbf k}  + \left(\epsilon_{\mathbf k} - E\right)
    {\tilde \alpha}^\dagger_{\mathbf k} {\tilde \alpha}_{\mathbf k} + \frac{|\Delta_0|^2}{2\beta \omega_0}\right]\nonumber \\
    &+&
  \frac{g}{\sqrt{2N\omega_0}}
   \sum_{{\mathbf k},{\mathbf q}} \left(\alpha_{\mathbf k}^\dagger \alpha_{{\mathbf k}+{\mathbf q}} + {\tilde \alpha}^\dagger_{\mathbf k} {\tilde \alpha}_{{\mathbf k}+{\mathbf q}} \right) (b^\dagger_{\mathbf q} + b_{-{\mathbf q}}).
   \label{qq_9_b_1}
   \eea
  where, as before,  $E =  \sqrt{{\mu_n}^2 + |\Delta_0|^2}$.
   We treat both $\Delta_0$ and $\mu_n$ as unknown parameters and obtain then from the condition that  density of $c_k$ fermions is $n$, and from the self-consistency condition on $\Delta_0$.  This leads to two equations
\bea
&& \frac{{\mu_n}}{E} \left[\frac{1}{N} \sum_{\mathbf k} \left( \langle \alpha^\dagger_{\mathbf k} \alpha_{\mathbf k} \rangle-  \langle {\tilde \alpha}^\dagger_{\mathbf k} {\tilde \alpha}_{\mathbf k}\rangle \right)\right]  =1-2n \nonumber \\
&&      1= - \frac{\beta \omega_0}{E} \frac{1}{N} \left[ \sum_{\mathbf k} \left(\langle \alpha^\dagger_{\mathbf k} \alpha_{\mathbf k} \rangle -
      \langle{\tilde \alpha}^\dagger_{\mathbf k} {\tilde \alpha}_{\mathbf k} \rangle \right)\right]
     \label{d_24_3_1}
     \eea
 Solving these equations for $W < 2\beta \omega_0$, where $ \langle \alpha^\dagger_{\mathbf k} \alpha_{\mathbf k} \rangle=0$ and
  $\langle {\tilde \alpha}^\dagger_{\mathbf k} {\tilde \alpha}_{\mathbf k} \rangle =1$,  we obtain $\mu_n = -\beta \omega_0 (1-2n)$ and $|\Delta_0| = 2\beta \omega_0 (n(1-n))^{1/2}$, like at $W=0$.

The Green's function of the physical $c$ fermion is a weighted sum of $G's$
 for  $\alpha$ and ${\tilde \alpha}$ fermions:
   \beq
   G^{c} (\omega, \epsilon_{\mathbf k}) =    \frac{E-{\mu_n}}{2E}\, G^{\alpha} (\omega, \epsilon_{\mathbf k}) +
   \frac{E+{\mu_n}}{2E}\,
    G^{{\tilde \alpha}} (\omega,\epsilon_{\mathbf k}).
  \label{qq_7_1_a_1}
   \eeq
Because
 $H^{'}_{\alpha}$  and $H^{'}_{{\tilde \alpha}}$ are
  the same as at $n=0$ and $n=1$, respectively, the evolution of the DOS with increasing $W$ proceeds in the same way as at $n=0$  for $\omega >0$ and at $n=1$ for $\omega <0$ (see Fig \ref{fig:extra} top in the main text).
  Like before, at $\sqrt{\beta} \omega_0 \ll W \ll \beta \omega_0$,
  the DOS consists of the polaron patches centered at  $|\omega| = l \omega_0$, where an integer $l \ll \beta$, and the two continua centered at $\omega = \pm \beta \omega_0$.
  The two continua touch each other  at $\lambda_0  = \lambda_0^{c,3}  = W N_F/2$, and start overlapping at smaller $\lambda_0$.  The DOS for these $\lambda_0$ is still different
  from that in a FL because  $\Delta_0$ is
   still finite.
   Solving Eqs.   (\ref{d_24_3_1}) for these $\lambda_0$,
  we find that $\mu_0 = -\beta \omega_0 (1-2n)$  as at larger $\lambda_0$,
   but $|\Delta_0|$ decreases and vanishes at $n-$dependent  $\lambda_0^{c,1} <W N_F/2$ (see Fig.
   \ref{fig:delta_vs_lambda} in the main text).  The Green's function of a physical $c$ fermion  at $\lambda_0^{c,1} < \lambda < \lambda_0^{c,3}$
    is
   \beq
   G^{c}  (\omega, \epsilon_{\mathbf k}) = \frac 12 \frac{1 - \mu_n/E}{\omega - (\epsilon_{\mathbf k} + E )} + \frac 12  \frac{1 + \mu_n /E}{\omega - (\epsilon_{\mathbf k} - E  )}
  \label{qq_7_1_a_2}
   \eeq
  In this range of $\lambda_0$, the DOS displays a pseudogap behavior (see Fig \ref{fig:extra} bottom in the main text).

   We verified that this a  pseudogap state can equally be viewed as a mixed state, in which a porion of the system with density $n_p = n-\delta$  remains  in
    the polaron state,   while a portion with density $\delta$ is in a FL state.
      The value of $\delta$  is obtained by minimizing  the sum of energies of the polaron and FL states.    For equilibrium $\delta$,  the sum of the two energies is exactly the same as the energy of the pseudogap state.   Furthermore,  $E = (\beta \omega_0 (1-2n)^2 + \Delta^2_0)^{1/2}$  with $\lambda-$dependent  $\Delta_0$  is exactly opposite to  the chemical potential of a FL with density $\delta$: $E = \mu_{FG} (\delta) = \beta \omega_0 (1-2 \delta)$. Due to this equivalence,  the Green's function  $G^{c}  (\omega, \epsilon_{\mathbf k}) $ from
      (\ref{qq_7_1_a_2}) can be re-expresses as the sum of the Green's function of free fermions with $\mu^{FL} (\delta)$  and the polaron part, for which   the DOS vanishes as $\omega =0$.  
The parameter $\delta$ varies between $\delta =0$ at $\lambda_0 = \lambda_0^{c,3} =  W N_F/2$ and $\delta =n$ at $\lambda_0 = \lambda_0^{c,1}$.  

The  area of the Fermi surface in the mixed state is $S_{FS} = \delta$ (the density of the FL component),  as one  can also obtain directly from
    (\ref{qq_7_1_a_2}).  It varies within the mixed state from $S_{FS} =0$ at the upper edge to $S_{FS} =n$ at the lower edge, where $\Delta_0 =0$.  The  Luttinger relation for an ordinary FL, $S_{FS} =n$,  is broken  because a portion of fermions with density $n_p = n - \delta$ form  localized polarons.
   We note in this regard that $G^{c}  (\omega, \epsilon_{\mathbf k})$ in
    (\ref{qq_7_1_a_2})
     has both poles and zeros, like in other cases, where the Luttinger theorem is broken
      \cite{Altshuler1997,Blason_2023,*Staffieri_2025,Lehmann_2025,*Stepanov_2024,Bonetti_2025,Bollmann2024,Kiran2013,Phillips2025}.

\end{document}